\documentclass[11pt,twoside]{article}
\usepackage{asp2010}

\resetcounters
\aspvolume{455}
\aspcpryear{2012}
\aspvoltitle{4$^{th}$ {\em Hinode} Science Meeting: Unsolved 
Problems and Recent Insights}
\aspvolauthor{L.~R.~Bellot Rubio, F.~Reale, and M.~Carlsson, eds.}

\begin{document}

\title{A Sharp Look at Coronal Rain with \textit{Hinode}/SOT and 
SST/CRISP}
\author{P.~Antolin,$^1$ M.~Carlsson,$^1$ L.~Rouppe van der Voort,$^1$ E.~Verwichte,$^2$ and G.~Vissers$^1$ 
\affil{$^1$Institute of Theoretical Astrophysics, University of Oslo, P.O. Box 1029, Blindern, NO-0315 Oslo, Norway}
\affil{$^2$Centre for Fusion, Space and Astrophysics, Department of Physics, University of Warwick, Coventry CV4 7AL, UK}}

\begin{abstract}
The tropical wisdom that when it is hot and dense we can expect rain
might also apply to the Sun. Indeed, observations and numerical
simulations have showed that strong heating at footpoints of loops, as
is the case for active regions, puts their coronae out of thermal
equilibrium, which can lead to a phenomenon known as catastrophic
cooling. Following local pressure loss in the corona, hot plasma
locally condenses in these loops and dramatically cools down to
chromospheric temperatures. These blobs become bright in H$\alpha$ and
\ion{Ca}{ii}~H in time scales of minutes, and their dynamics seem to
be subject more to internal pressure changes in the loop rather than
to gravity. They thus become trackers of the magnetic field, which
results in the spectacular coronal rain that is observed falling down
coronal loops. In this work we report on high resolution observations
of coronal rain with the Solar Optical Telescope (SOT) on
\textit{Hinode} and CRISP at the Swedish Solar Telescope
(SST). A statistical study is performed in which properties
such as velocities and accelerations of coronal rain are derived. We
show how this phenomenon can constitute a diagnostic tool for the
internal physical conditions inside loops. Furthermore, we analyze
transverse oscillations of strand-like condensations composing coronal
rain falling in a loop, and discuss the possible nature of the
wave. This points to the important role that coronal rain can play in
the fields of coronal heating and coronal seismology.

\end{abstract}

\section{Introduction}

There is now increasing evidence that active regions in the Sun have
their heating concentrated mostly at lower atmospheric regions, from
the lower chromosphere to the lower corona. The excess densities found
in most observed coronal structures such as coronal loops put them out
of hydrostatic equilibrium, a state that can be explained by basal
heating \citep{2001ApJ...560.1035A}. \citet{2008ApJ...678L..67H},
using the \textit{Hinode}/EIS instrument, have shown that active
region loops exhibit upflow motions and enhanced nonthermal velocities
at their footpoints, fitting as well in the basal heating
scenario. Recently, \citet{2011Sci...331...55D} have highlighted the
importance of the link between the photosphere and the corona by
showing that a considerable part of the hot coronal plasma gets heated
at low spicular heights, thus explaining the fading character of the
ubiquitous ``type II spicules''. Further evidence of basal heating is
put forward by the presence of cool structures in the active region
coronae, such as filaments/prominences or coronal rain, two phenomena
that are linked to the underlying magnetic field topology.

Both prominences (or filaments if observed on disk rather than at the
limb) and coronal rain correspond to cool and dense plasma observed at
coronal heights in chromospheric lines such as H$\alpha$ and
\ion{Ca}{ii}~H and K. But while the plasma in prominences is suspended
in the corona against gravity making the structures long-lived (days
to weeks), coronal rain is observed falling down in timescales of
minutes \citep{2004A&A...415.1141D, 2001SoPh..198..325S} along curved
loop-like trajectories in the absence of prominences. It is generally
agreed that the mechanical stability and thermodynamic properties of
prominences are deeply linked to the underlying magnetic field
topology, and thus we could think that the main difference with
coronal rain is basically a difference on the magnetic field
configuration in the corona. This is, however, a speculation that
still needs to be addressed properly.

While prominences have been studied extensively in solar physics,
coronal rain, on the other hand, is a phenomenon for which few
observational studies exist, despite being observed since the early
1970s \citep{1970PASJ...22..405K, 1972SoPh...25..413L}. It is also
generally believed to be a rather uncommon phenomenon of active region
coronae. However, recent high resolution observations with instruments
such as the CRisp Imaging Spectro-Polarimeter (CRISP) at the Solar
Swedish Telescope (SST), the Solar Optical Telescope on
\textit{Hinode}, or the Solar Dynamics Observatory (SDO) are
offering a different scenario in which coronal rain appears to be a
rather ubiquitous phenomenon over active regions, a case which can
have profound implications in coronal heating. Whether this is really
the case will require further effort gathering a statistically
significant number of observations in different wavelengths.

Numerical simulations have shown that coronal rain and prominences are most likely the result of a phenomenon of thermal instability, also known as 'catastrophic cooling'. Loops with basal heating present high coronal densities and thermal conduction turns out to be insufficient to maintain a steady heating per unit mass, leading to a gradual decrease of the coronal temperature. Eventually recombination of atoms takes place and temperature decreases to chromospheric values abruptly in a timescale of minutes locally in the corona. This is accompanied by local pressure losses leading to the formation of condensations which become bright or dark if observed towards the limb or on disk, respectively.

The catastrophic cooling leading to coronal rain is not due to a lack
of heating flux into the loop, neither to the temporal characteristics
of the heating, as shown by \citet{1999ApJ...512..985A}, 
\citet{2002ApJ...579L..49M}, \citet{2003A&A...411..605M}, and 
\citet{2004A&A...424..289M}. Rather it is linked to a specific 
spatial distribution of the heating, namely, basal heating. Only
heating mechanisms having such heating scale heights allow
catastrophic cooling to happen, thus pointing to the important link
between coronal rain and coronal heating. Along these lines,
\citet{2010ApJ...716..154A} showed that if Alfv\'en wave heating is
predominant in the loop, coronal rain is inhibited due to the
characteristic uniform spatial heating ensuing from Alfv\'en waves.

High resolution observations in the last decade have unveiled a new
important aspect of prominences: their thread-like structure
\citep{2006ApJ...643L..65H, 2005SoPh..226..239L, 2008ASPC..383..235L,
2011SSRv..158..237L}. At these ultra-small scales, prominences are
observed to be composed of a myriad of fine threads (widths on the
order of 200~km or less), outlining a fine-scale structure of the
magnetic field. This magnetic field correspondence is further
evidenced by the presence of flows along these threads
\citep{2008SoPh..250...31M}. Although the prominence as a whole is a
long-lived and rather static structure, the observed thread-like
components appear to be very dynamic and short-lived (1--10
minutes). Likewise, very high resolution observations of coronal rain
with \textit{Hinode}/SOT have shown strand-like structures with widths
below 200 km \citep{2010ApJ...716..154A} traced by the condensations
separating and elongating during their fall.
 
The detection of coronal rain requires therefore very high resolution
observations. In this work we present such observations and perform a
statistical study of velocities, accelerations, and falling angles of
coronal rain observed with CRISP at the SST and SOT on
\textit{Hinode}. The CRISP spectropolarimeter allows us to obtain a
more precise picture on the dynamics since it provides us with Doppler
information, additionally to imaging. The obtained velocities are well
below free fall, as has been reported in the past for prominences as
well \citep{2001SoPh..198..325S, 1994SoPh..150...81Z,
2001SoPh..198..289M}. This accounts for the presence of other forces
than gravity inside loops, most likely of magnetic origin, which may
be related to the forces supporting prominences. The shapes of the
falling condensations offer further evidence of very thin thread-like
structures. Dynamics and shapes offer constraints on the nature of the
forces inside the loops to which coronal rain is subject.

Waves have been observed frequently along prominences
\citep{1966AJ.....71..197R, 2002SoPh..206...45O, 2007Sci...318.1577O,
2011SSRv..158..237L} and coronal loops \citep{2005LRSP....2....3N,
2007SoPh..246....3B, 2009SSRv..149..199R, 2009SSRv..149..229T},
leading to the determination of the internal physical conditions
through the development of analytical and numerical modeling, thus
spawning a very active research community in the field of prominence
seismology. In this work, we also present the first observational
analysis of transverse oscillations of threads in a loop, put in
evidence by coronal rain. Strand-like condensations composing coronal
rain are found to oscillate in phase, indicating transverse waves,
probably of a collective nature. Periods, amplitudes, transversal
velocities and phase speeds of the waves, as well as the coronal
magnetic field, are estimated. We illustrate through this observational
result the potential that coronal rain can have in the field of
coronal seismology.

\section{Observations with SST/CRISP and \textit{Hinode}/SOT}

\begin{figure}[!t]
\centering
\includegraphics[scale=1.23,bb= 0 15 226 192,clip]{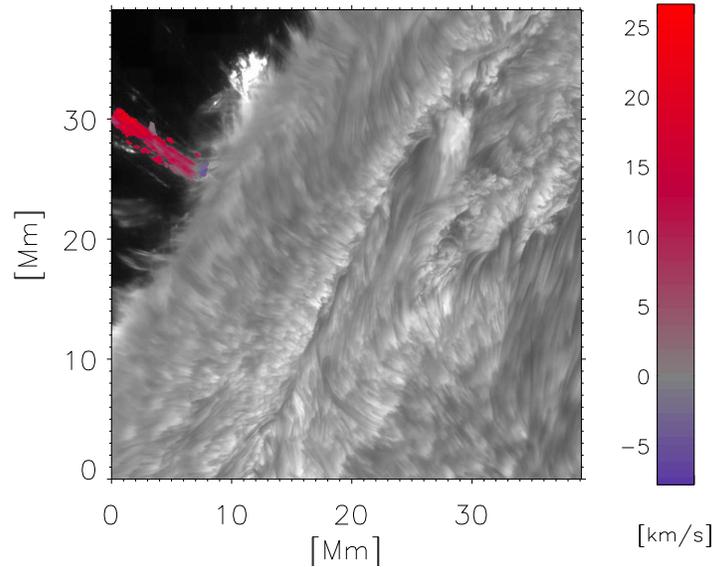}
\caption{Coronal rain observed with CRISP in H$\alpha+0.03$~nm 
on May 10, 2009 above AR 11017. We plot in color the Doppler
velocities for a loop with coronal rain.}
\label{fig1}
\end{figure}

The observations with CRISP \citep{2008ApJ...689L..69S} at the SST
\citep{2003SPIE.4853..341S} were done in
the spectral window [$-0.12$, 0.12]~nm centered in H$\alpha$, on May 10,
2009 from 08:50 to 09:37 UT with a cadence of 6.4~s and a resolution
of 0\farcs0592~pixel$^{-1}$, focusing on the active
region (AR) 11017 on the east solar limb. The observations with 
\textit{Hinode}/SOT \citep{2008SoPh..249..167T} were done in 
the \ion{Ca}{ii}~H band on November 9, 2006 from 19:33 to 20:44 UT
with a cadence of 15~s and a resolution of 0\farcs05448~pixel$^{-1}$,
and focused on NOAA AR 10921 on the west limb. A statistical study
focusing on the coronal rain observations by \textit{Hinode}/SOT has
already been done in
\citet{2010ApJ...716..154A}. The statistical study presented here
concentrates on the observations by CRISP.

In Fig. \ref{fig1} we present a snapshot of the coronal rain falling
down along a system of loops observed by CRISP of SST in
$H\alpha+0.03$ nm. In color we plot the Doppler velocities derived by
the spectropolarimeter over some of the loops exhibiting coronal
rain. A radial filter which enhances the intensity of features above
the limb has also been applied to the image. By tracing the falling
condensations composing coronal rain we have calculated the velocities
(projected and Doppler velocities), accelerations, and falling angles
(with respect to the vertical) for a total of 235 condensations
tracing 22 strands in total. This is shown in the histogram panels of
Figure~\ref{fig2}. As the condensations fall in general with
non-constant velocities, we have measured the velocity for each
condensation at two different heights, one high in the corona and
another one towards the footpoints in the chromosphere, thus leading
to an estimate of their acceleration, which we plot in the middle
panel of the figure. However, only 76 out of the 235 condensations
could be followed clearly along the strands. The rest of the
condensations can mostly be discerned towards the footpoints. In the
velocity panel on the left of Fig.~\ref{fig2} we have considered all
the measurements, irrespective of their location along the
strands. The velocities correspond to the total velocities, namely,
the magnitude of the vector sum of the projected velocities in the
plane of the sky (perpendicular to the line of sight) and the Doppler
velocities. From these two components of the velocity we can calculate
the angle at which the condensations fall with respect to the
vertical. The histogram for the calculated angles is shown in the
right panel of Figure~\ref{fig2}. In all 3 panels, the dashed line
corresponds to the mean values, $70\pm23$~km~s$^{-1}$,
$0.12\pm0.17$~km~s$^{-2}$, and $13\fdg4 \pm 10\deg$, respectively,
with their standard deviations.

\begin{table}[t]
\begin{center}
\caption{Mean periods, amplitudes and transversal velocities 
for the 9 oscillatory events detected by \textit{Hinode}/SOT on
November 9, 2006 in a loop with coronal rain. Phase speeds and coronal
magnetic fields are estimated according to the text.}
\begin{tabular}{lccccc}
\noalign{\smallskip}
\hline
Event & Period & Amplitude & Transversal & Phase speed & Magnetic
field \\ & [s] & [Mm] & velocity [km~s$^{-1}$] & [km~s$^{-1}$] & [G] \\
\hline
1 & 112 $\pm$ 29 & 245 $\pm$ 148 &  4.5 $\pm$ 2.5 & 760 $\pm$ 265 & 16 $\pm$ 5.5\\
2 & 171 $\pm$ 11 & 515 $\pm$ 135 &  6.1 $\pm$ 1.4 & 460 $\pm$ 30 & 9.7 $\pm$ 0.6\\
3 & 118 $\pm$ 15 & 308 $\pm$ 136 &  5.3 $\pm$ 1.3 & 675 $\pm$ 90 & 14.2 $\pm$ 1.9\\
4 & 143 $\pm$ 17 & 324 $\pm$ 125 &  4.4 $\pm$ 1.2 & 555 $\pm$ 65 & 11.6 $\pm$ 1.4\\
5 & 165 $\pm$ - & 351 $\pm$ 55 &  4.2 $\pm$ 0.1 & 475 $\pm$ - & 10 $\pm$ - \\
6 & 198 $\pm$ 49 & 371 $\pm$ 205 &  3.6 $\pm$ 2.0 & 410 $\pm$ 90 & 8.6 $\pm$ 1.9\\
7 & 172 $\pm$ 20 & 406 $\pm$ 179 &  4.2 $\pm$ 1.0 & 460 $\pm$ 60 & 9.6 $\pm$ 1.2\\
8 & 176 $\pm$ 23 & 369 $\pm$ 255 &  3.8 $\pm$ 2.7 & 450 $\pm$ 60 & 9.5 $\pm$ 1.2\\
9 & 84 $\pm$ 8 & 305 $\pm$ 119 & 7.6 $\pm$ 4.0 & 940 $\pm$ 95 & 19.7 $\pm$ 2 \\
\hline
\end{tabular}
\end{center}
\label{tab1}
\end{table}

In Fig.~\ref{fig3} we present the observations by
\textit{Hinode}/SOT. In the left panel we show a subset of the entire
field of view, in which we plot the variance of the image over the
period of time when coronal rain falls along the loop (about 15
minutes). We can see that the plane of the loop is roughly directed
along the line of sight. The coronal rain in this case can be observed
basically from the apex of the loop, located $25\pm5$~Mm above the
surface, leading to a loop length of $80\pm15$~Mm assuming a circular
loop. The dynamics of the condensations are similar to those
observed by the SST and reported here \citep{2010ApJ...716..154A}. The
condensations separate and elongate, thus tracing several strands in
the loop, and put in evidence in-phase transverse oscillations of the
strands. This is shown in the right panel of Figure~\ref{fig3}. Here
we show the time slice along the outlined loop, where the transverse
length refers to the perpendicular distance to the dotted line from
dashed line to dashed line. At least 9 oscillation patterns can be
clearly observed, over which we plot in color the (projected) distance
where they happen from the apex of the loop. The in-phase oscillation
can be observed in strands 1, 2, 3, 4, and 6 between the time period [15,
22] min in the figure. These strands become all visible simultaneously
at roughly 8 Mm from the apex. The estimated periods, amplitudes,
transversal velocities are shown in Table \ref{tab1}, as well as
estimations for phase speeds and coronal magnetic fields according to
the wave interpretation (see the next section).

\section{Discussion}

\subsection{Statistics of Coronal Rain from CRISP Observations}

The coronal rain observed at the limb with the SST falls down roughly
perpendicular to the solar surface. As shown in Fig.~\ref{fig1}, CRISP
indicates that most of the condensations are redshifted. Furthermore,
the Doppler velocity of the condensations is observed to decrease with
height above the solar surface. These facts fit in a scenario in which
the loops have their planes roughly directed along the line of sight,
with their farthest legs present in the field of view of the SST.

\begin{figure}[t]
\begin{center}
\includegraphics[scale=0.318,bb=20 10 407 360]{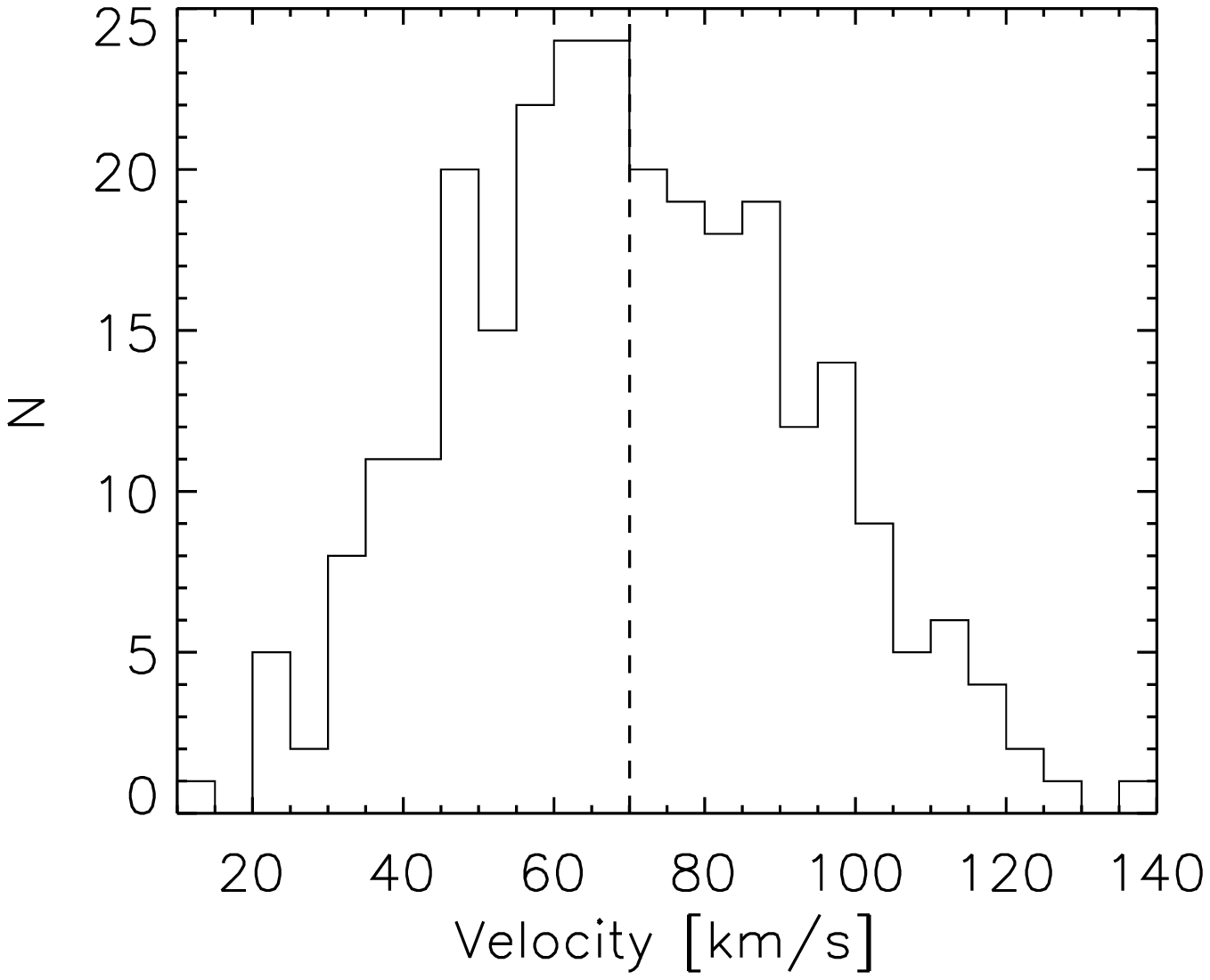}   
\includegraphics[scale=0.318,bb=20 10 407 360]{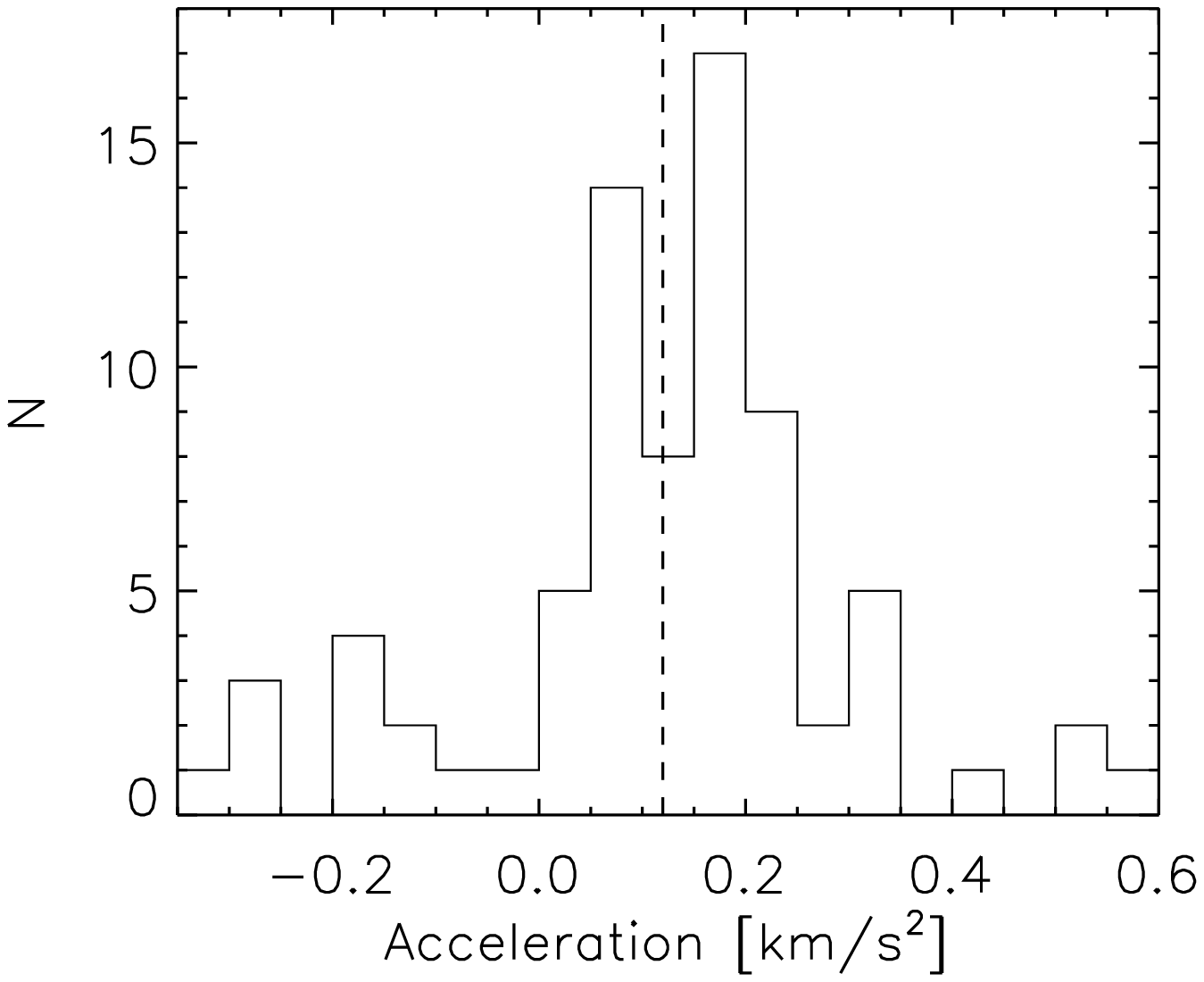} 
\includegraphics[scale=0.318,bb=20 10 407 360]{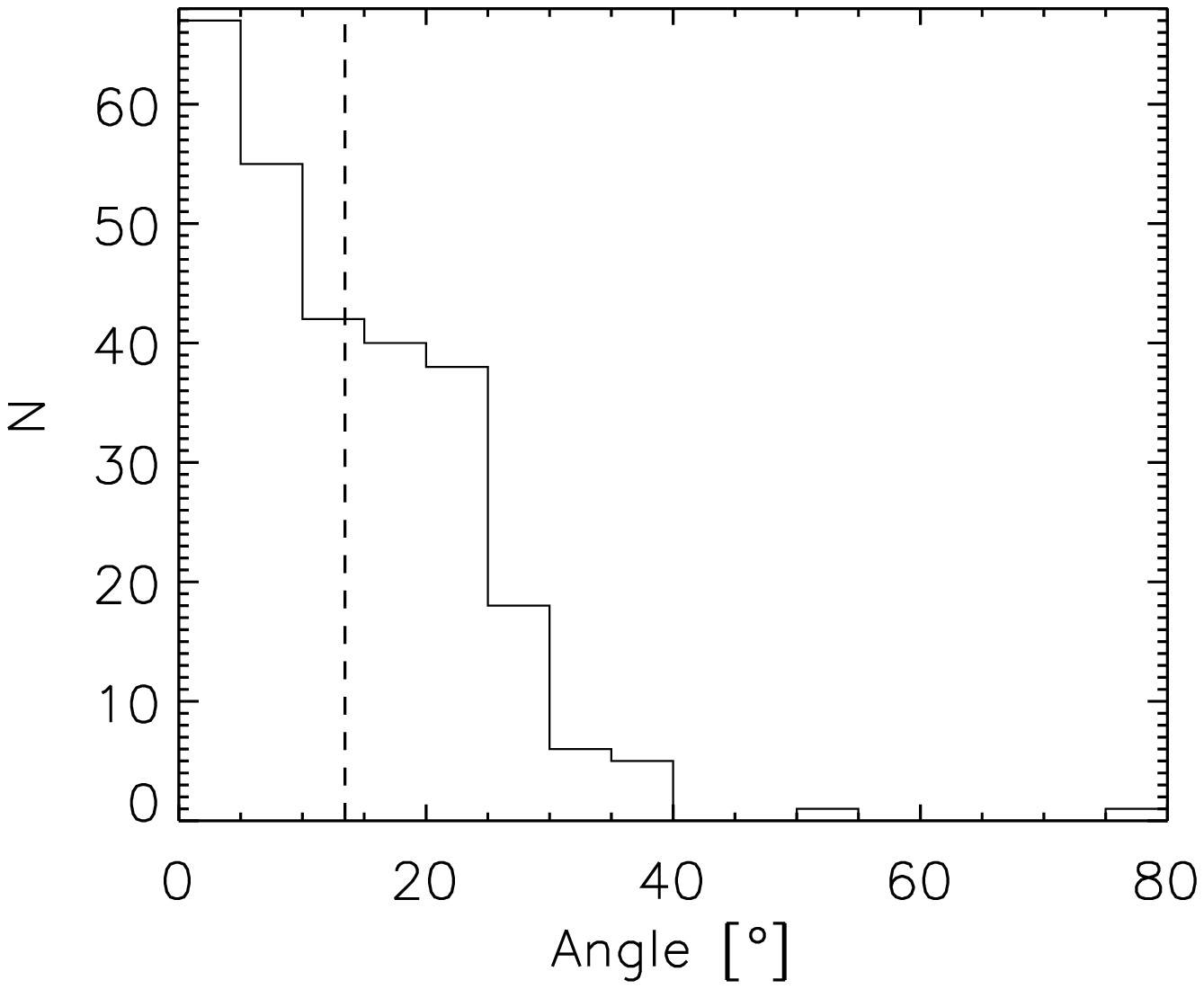}   
\caption{Histograms of velocity (\textit{left}), acceleration (\textit{middle}) and falling angle with respect to vertical (\textit{right}) for the coronal rain observed with CRISP in Figure~\ref{fig1}.}
\label{fig2}
\end{center}
\end{figure}

The observed coronal rain has a broad distribution of velocities, as shown in the left panel of Figure~\ref{fig2}. This broadness comes not from a uniform acceleration of initially slowly falling condensations under the action of an effective gravity along the loops, but is the result of varying acceleration and deceleration processes. Indeed, the histogram in the middle panel of the figure shows an important spreading of accelerations, reaching even negative values indicating deceleration processes. As the mean of the distribution of the accelerations indicates, $0.12\pm0.17$ km s$^{-2}$, almost all values are below the solar gravity value. Assuming nonetheless that the effective gravity (component along the loop) plays the main role in the dynamics of the condensations, the obtained mean value for the acceleration would be the result of falling at a mean angle of 65$\deg$. However, the inferred falling angles from the observed velocity components are almost all below 30$\deg$, as shown in the right panel of the figure. In fact, most of the observed heights for the condensations are between 10 and 20 Mm. Correlating the observed heights with the falling angles and assuming circular loops we obtain a best fit for the heights of the loop apexes of $38\pm13$ Mm, leading to total lengths of $120\pm40$ Mm, which indicates that the observed part of the loops corresponds only to the lower legs. 

In both SST and \textit{Hinode} observations the condensations composing coronal rain are often observed to separate as they fall, resulting in elongation processes leading to thin thread-like structures of 200 km or lower. Thus, coronal rain in high resolution images is often observed to track strand-like structures within coronal loops. Whether the separation and elongation processes are just a result of gravity acting differentially along the magnetic topology or if other more sophisticated processes are involved is an interesting question that needs to be addressed properly with the help of numerical simulations. For instance, since the condensations appear to have chromospheric-like densities, thus becoming optically thicker, the local plasma beta parameter could be high enough to allow the plasma in the condensation to move transversally to the loop axis, thus allowing the separation of the initial dense blob.

\subsection{Transverse MHD waves in coronal loops highlighted by coronal rain}

\begin{figure}
\begin{center}
$\begin{array}{c@{\hspace{-0.25in}}c}
\includegraphics[scale=0.44]{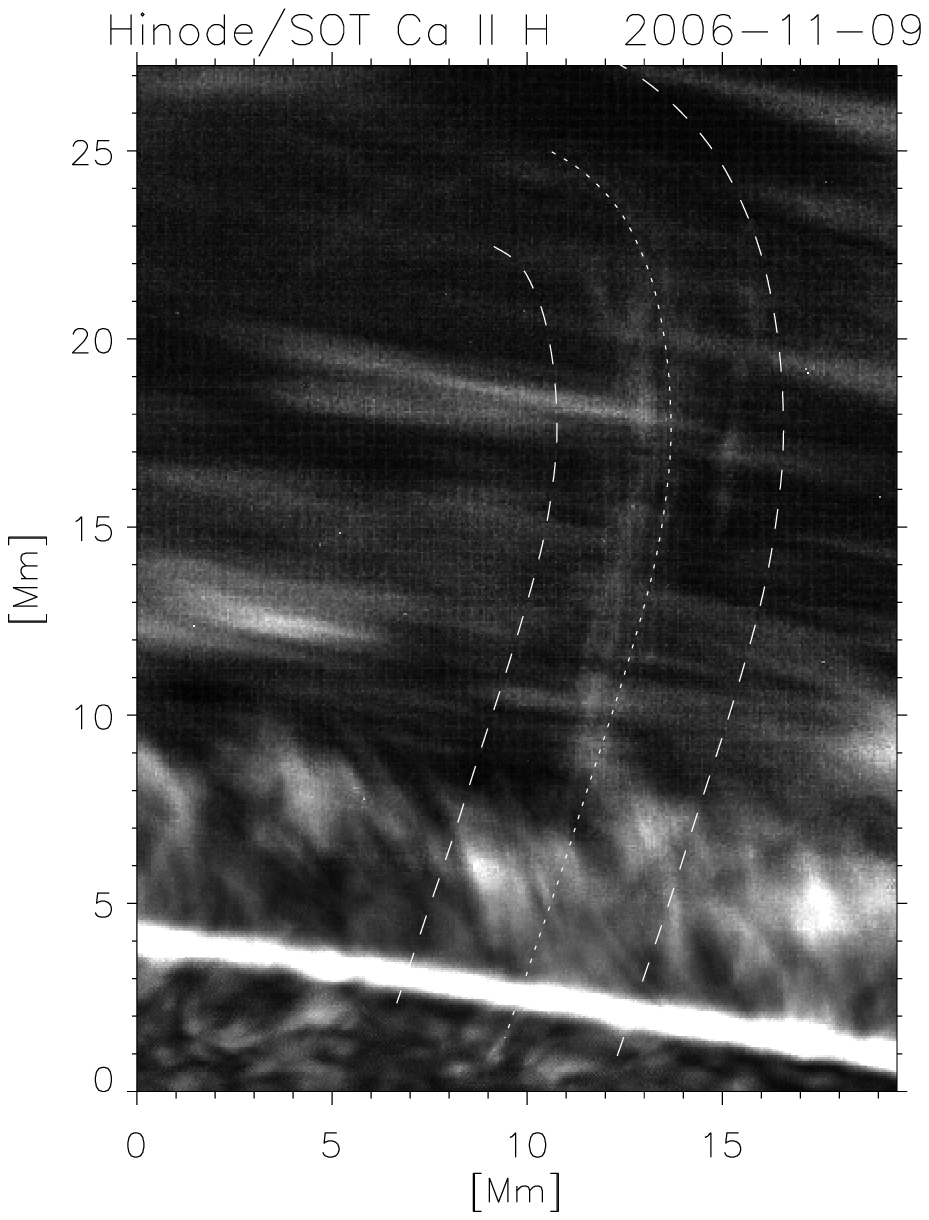}  &  
\includegraphics[scale=0.44]{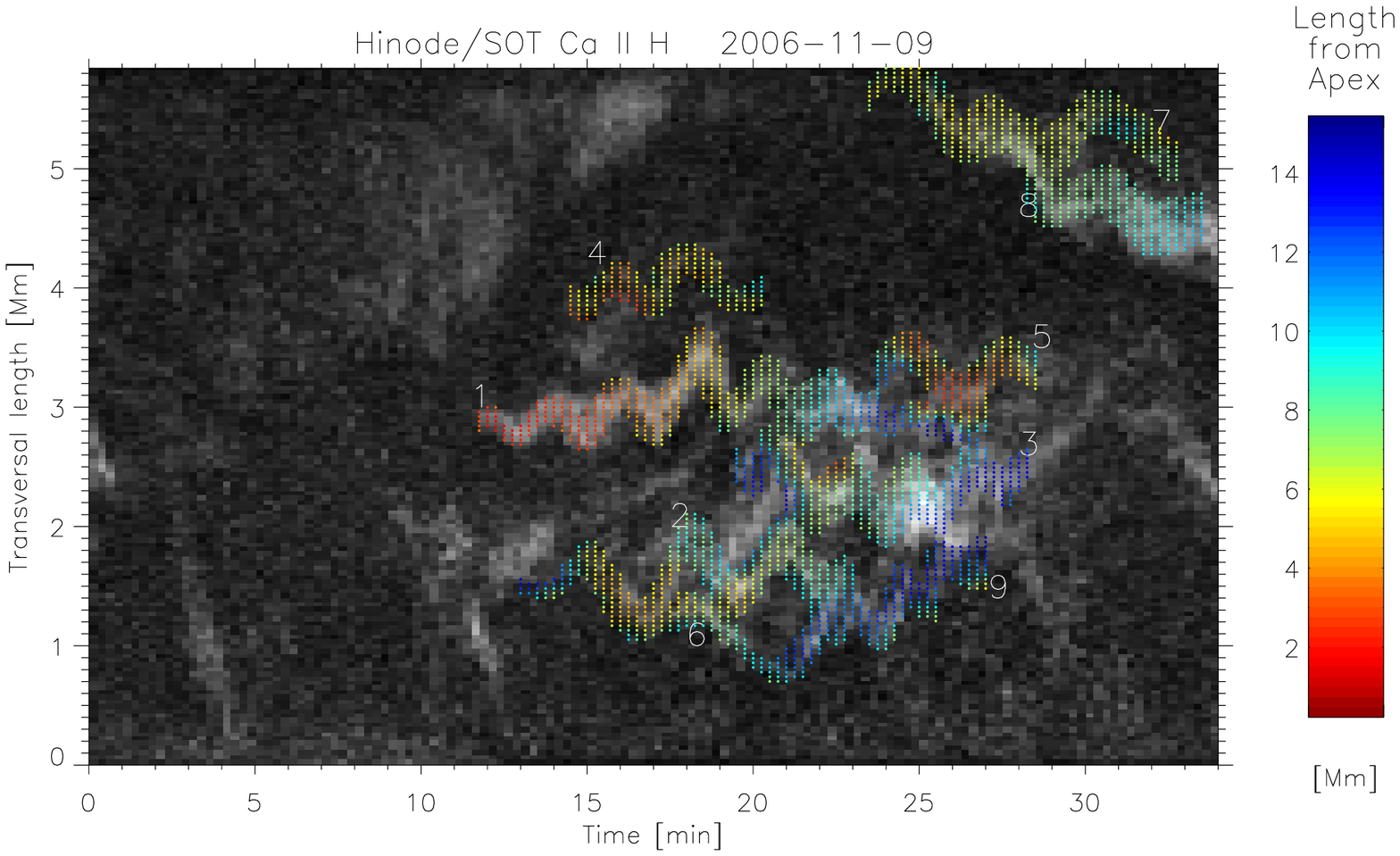}
\end{array}$
\caption{\textit{Left:} Loop with coronal rain observed by 
\textit{Hinode}/SOT on November 9, 2006 above AR 10921. \textit{Right:} 
Time slice across the loop (transversal width corresponds to width of
the loop delimited by the dashed lines in the left panel) for the time
interval when coronal rain is observed. 9 oscillations can be
detected. We plot in color over the oscillations the length from the
apex where they occur.}
\vspace*{-1em}
\label{fig3}
\end{center}
\end{figure}

The data set obtained by \textit{Hinode}/SOT has become famous among
solar limb observations. This is due mainly to the presence of an
active region prominence which exhibits interesting oscillatory
behavior. The waves running through this prominence were first
interpreted as Alfv\'en waves \citep{2007Sci...318.1577O}, although
various arguments have been put forward for an interpretation in terms
of fast kink modes \citep{2008ApJ...676L..73V,
2008ApJ...678L.153T}. The reported mean periods for the waves are
between 130 and 240 s, (horizontal) oscillation amplitudes between 400
and 1770 km, transverse (vertical) velocities between 5 and 15 km
s$^{-1}$ and an estimated wave speed of $>1050$ km s$^{-1}$ leading to
a magnetic field of $50$ G in the prominence.

Additionally, on the foreground of the prominence various loops exhibiting coronal rain have been observed \citep{2010ApJ...716..154A}. In one of these loops, located on the south side of the \textit{Hinode}/SOT field of view (right hand side of Figure~1 in that paper, left panel in Figure~\ref{fig3} here) transverse oscillations are put in evidence by the coronal rain. Several strand-like condensations are observed to oscillate in-phase, which points to a transverse wave affecting the loop as one structure, rather than being random local perturbation processes in the strands. The oscillations can not only be confined to the loop but can involve a larger coronal region, and thus could be related to the oscillations reported by \citet{2007Sci...318.1577O} present in the background prominence. The periods are similar but the amplitudes are somewhat smaller in our case, and more importantly, the oscillation is along the horizontal plane, whereas the oscillation in the prominence is along the vertical plane. Furthermore, the loops on the north side of the sunspot \citep[analyzed in][]{2010ApJ...716..154A} do not exhibit oscillatory motions.

Since the oscillations in the loop can only be observed when the
condensations are falling it is not clear whether the agent is a
propagating or a standing wave. The condensations do not appear to be
oscillating when they are at the apex of the loop (as shown for
instance by strand 1 in Fig.~\ref{fig3}) and the amplitudes indicate
a maximum at roughly halfway along the loop leg (one fourth of the
total loop length), which correspond to signatures of the first
harmonic of a standing mode. On the other hand, we could think of a
propagating wave packet, propagating up or down, the maximum amplitude
meeting the condensations half way through the loop's leg. Although
the latter scenario is less likely, it cannot be excluded.

In the case of a first harmonic, the total wavelength is equal to the loop length, i.e. $\simeq80$~Mm, leading to phase speeds between $400$ and 1000 km s$^{-1}$ (see Table \ref{tab1}). If we assume that the density inside the loop is quite higher than the density outside the loop (as expected from catastrophic cooling loops), we would be assuming the presence of a waveguide along the loop, in which case the wave causing the oscillations would be a fast kink wave. Taking a typical density ratio of 0.1 between outside and inside densities, assuming a loop number density of $3\times10^{9}$ cm$^{-3}$, a rough maximum limit of dense loops subject to catastrophic cooling \citep{2010ApJ...716..154A} and using formula 31 in \citet{2005LRSP....2....3N} (corrected for the first harmonic), we obtain a coronal magnetic field in the loop between 9 and 20~G. It is important to note, however, that the condensations do not oscillate in-phase throughout the entire falling time. The loss of phase may be a signature of a non-collective character of the wave. A torsional Alfv\'en wave is known to exhibit such behavior due to phase mixing.

The causes of such oscillation are less clear. The most natural
scenario is one in which the waves are generated at photospheric level
by magnetic reconnection. We can also think of a special case in which
coronal rain itself leads to a kink mode in the loop, a scenario worthy
of investigation through numerical simulations.

\acknowledgements P.A. would like to thank the SOC and LOC for the
opportunity to present this work.  

\bibliography{antolin.bbl} 

\begin{thebibliography}{}
\expandafter\ifx\csname natexlab\endcsname\relax\def\natexlab#1{#1}\fi
\expandafter\ifx\csname url\endcsname\relax
  \def\url#1{\texttt{#1}}\fi
\expandafter\ifx\csname urlprefix\endcsname\relax\def\urlprefix{URL }\fi
\providecommand{\eprint}[2][]{\url{#2}}

\bibitem[{{Antiochos} et~al.(1999){Antiochos}, {MacNeice}, {Spicer}, \&
  {Klimchuk}}]{1999ApJ...512..985A}
{Antiochos}, S.~K., {MacNeice}, P.~J., {Spicer}, D.~S., \& {Klimchuk}, J.~A.
  1999, \apj, 512, 985

\bibitem[{{Antolin} et~al.(2010){Antolin}, {Shibata}, \&
  {Vissers}}]{2010ApJ...716..154A}
{Antolin}, P., {Shibata}, K., \& {Vissers}, G. 2010, \apj, 716, 154

\bibitem[{{Aschwanden}(2001)}]{2001ApJ...560.1035A}
{Aschwanden}, M.~J. 2001, \apj, 560, 1035

\bibitem[{{Banerjee} et~al.(2007){Banerjee}, {Erd{\'e}lyi}, {Oliver}, \&
  {O'Shea}}]{2007SoPh..246....3B}
{Banerjee}, D., {Erd{\'e}lyi}, R., {Oliver}, R., \& {O'Shea}, E. 2007,
  \solphys, 246, 3

\bibitem[{{De Groof} et~al.(2004){De Groof}, {Berghmans}, {van
  Driel-Gesztelyi}, \& {Poedts}}]{2004A&A...415.1141D}
{De Groof}, A., {Berghmans}, D., {van Driel-Gesztelyi}, L., \& {Poedts}, S.
  2004, \aap, 415, 1141

\bibitem[{{De Pontieu} et~al.(2011){De Pontieu}, {McIntosh}, {Carlsson},
  {Hansteen}, {Tarbell}, {Boerner}, {Mart{\'{\i}nez}-Sykora}, {Schrijver}, \&
  {Title}}]{2011Sci...331...55D}
{De Pontieu}, B., {McIntosh}, S.~W., {Carlsson}, M., {Hansteen}, V.~H.,
  {Tarbell}, T.~D., {Boerner}, P., {Mart{\'{\i}nez}-Sykora}, J., {Schrijver},
  C.~J., \& {Title}, A.~M. 2011, Science, 331, 55

\bibitem[{{Hara} et~al.(2008){Hara}, {Watanabe}, {Harra}, {Culhane}, {Young},
  {Mariska}, \& {Doschek}}]{2008ApJ...678L..67H}
{Hara}, H., {Watanabe}, T., {Harra}, L.~K., {Culhane}, J.~L., {Young}, P.~R.,
  {Mariska}, J.~T., \& {Doschek}, G.~A. 2008, \apjl, 678, L67

\bibitem[{{Heinzel} \& {Anzer}(2006)}]{2006ApJ...643L..65H}
{Heinzel}, P., \& {Anzer}, U. 2006, \apjl, 643, L65

\bibitem[{{Kawaguchi}(1970)}]{1970PASJ...22..405K}
{Kawaguchi}, I. 1970, \pasj, 22, 405

\bibitem[{{Leroy}(1972)}]{1972SoPh...25..413L}
{Leroy}, J.-L. 1972, \solphys, 25, 413

\bibitem[{{Lin}(2010)}]{2011SSRv..158..237L}
{Lin}, Y. 2010, Space~Sci.~Rev., 112

\bibitem[{{Lin} et~al.(2005){Lin}, {Engvold}, {Rouppe van der Voort}, {Wiik},
  \& {Berger}}]{2005SoPh..226..239L}
{Lin}, Y., {Engvold}, O., {Rouppe van der Voort}, L., {Wiik}, J.~E., \&
  {Berger}, T.~E. 2005, \solphys, 226, 239

\bibitem[{{Lin} et~al.(2008){Lin}, {Martin}, \&
  {Engvold}}]{2008ASPC..383..235L}
{Lin}, Y., {Martin}, S.~F., \& {Engvold}, O. 2008, in Subsurface and
  Atmospheric Influences on Solar Activity, edited by {R.~Howe, R.~W.~Komm,
  K.~S.~Balasubramaniam, \& G.~J.~D.~Petrie }, vol. 383 of Astronomical Society
  of the Pacific Conference Series, 235

\bibitem[{{Mackay} \& {Galsgaard}(2001)}]{2001SoPh..198..289M}
{Mackay}, D.~H., \& {Galsgaard}, K. 2001, \solphys, 198, 289

\bibitem[{{Martin} et~al.(2008){Martin}, {Lin}, \&
  {Engvold}}]{2008SoPh..250...31M}
{Martin}, S.~F., {Lin}, Y., \& {Engvold}, O. 2008, \solphys, 250, 31

\bibitem[{{Mendoza-Brice{\~n}o} et~al.(2002){Mendoza-Brice{\~n}o},
  {Erd{\'e}lyi}, \& {Di G.~Sigalotti}}]{2002ApJ...579L..49M}
{Mendoza-Brice{\~n}o}, C.~A., {Erd{\'e}lyi}, R., \& {Di G.~Sigalotti}, L. 2002,
  \apjl, 579, L49

\bibitem[{{M{\"u}ller} et~al.(2003){M{\"u}ller}, {Hansteen}, \&
  {Peter}}]{2003A&A...411..605M}
{M{\"u}ller}, D.~A.~N., {Hansteen}, V.~H., \& {Peter}, H. 2003, \aap, 411, 605

\bibitem[{{M{\"u}ller} et~al.(2004){M{\"u}ller}, {Peter}, \&
  {Hansteen}}]{2004A&A...424..289M}
{M{\"u}ller}, D.~A.~N., {Peter}, H., \& {Hansteen}, V.~H. 2004, \aap, 424, 289

\bibitem[{{Nakariakov} \& {Verwichte}(2005)}]{2005LRSP....2....3N}
{Nakariakov}, V.~M., \& {Verwichte}, E. 2005, Living Reviews in Solar Physics,
  2, 3

\bibitem[{{Okamoto} et~al.(2007){Okamoto}, {Tsuneta}, {Berger}, {Ichimoto},
  {Katsukawa}, {Lites}, {Nagata}, {Shibata}, {Shimizu}, {Shine}, {Suematsu},
  {Tarbell}, \& {Title}}]{2007Sci...318.1577O}
{Okamoto}, T.~J., {Tsuneta}, S., {Berger}, T.~E., {Ichimoto}, K., {Katsukawa},
  Y., {Lites}, B.~W., {Nagata}, S., {Shibata}, K., {Shimizu}, T., {Shine},
  R.~A., {Suematsu}, Y., {Tarbell}, T.~D., \& {Title}, A.~M. 2007, Science,
  318, 1577

\bibitem[{{Oliver} \& {Ballester}(2002)}]{2002SoPh..206...45O}
{Oliver}, R., \& {Ballester}, J.~L. 2002, \solphys, 206, 45

\bibitem[{{Ramsey} \& {Smith}(1966)}]{1966AJ.....71..197R}
{Ramsey}, H.~E., \& {Smith}, S.~F. 1966, \aj, 71, 197

\bibitem[{{Ruderman} \& {Erd{\'e}lyi}(2009)}]{2009SSRv..149..199R}
{Ruderman}, M.~S., \& {Erd{\'e}lyi}, R. 2009, Space~Sci.~Rev., 149, 199

\bibitem[{{Scharmer} et~al.(2003){Scharmer}, {Bjelksjo}, {Korhonen},
  {Lindberg}, \& {Petterson}}]{2003SPIE.4853..341S}
{Scharmer}, G.~B., {Bjelksjo}, K., {Korhonen}, T.~K., {Lindberg}, B., \&
  {Petterson}, B. 2003, in Innovative Telescopes and Instrumentation for Solar
  Astrophysics, edited by {S.~L.~Keil \& S.~V.~Avakyan}, vol. 4853 of Society
  of Photo-Optical Instrumentation Engineers (SPIE) Conference Series, 341

\bibitem[{{Scharmer} et~al.(2008){Scharmer}, {Narayan}, {Hillberg}, {de la Cruz
  Rodr{\'{\i}}guez}, {L{\"o}fdahl}, {Kiselman}, {S{\"u}tterlin}, {van Noort},
  \& {Lagg}}]{2008ApJ...689L..69S}
{Scharmer}, G.~B., {Narayan}, G., {Hillberg}, T., {de la Cruz
  Rodr{\'{\i}}guez}, J., {L{\"o}fdahl}, M.~G., {Kiselman}, D., {S{\"u}tterlin},
  P., {van Noort}, M., \& {Lagg}, A. 2008, \apjl, 689, L69

\bibitem[{{Schrijver}(2001)}]{2001SoPh..198..325S}
{Schrijver}, C.~J. 2001, \solphys, 198, 325

\bibitem[{{Taroyan} \& {Erd{\'e}lyi}(2009)}]{2009SSRv..149..229T}
{Taroyan}, Y., \& {Erd{\'e}lyi}, R. 2009, Space~Sci.~Rev., 149, 229

\bibitem[{{Terradas} et~al.(2008){Terradas}, {Arregui}, {Oliver}, \&
  {Ballester}}]{2008ApJ...678L.153T}
{Terradas}, J., {Arregui}, I., {Oliver}, R., \& {Ballester}, J.~L. 2008, \apjl,
  678, L153

\bibitem[{{Tsuneta} et~al.(2008){Tsuneta}, {Ichimoto}, {Katsukawa}, {Nagata},
  {Otsubo}, {Shimizu}, {Suematsu}, {Nakagiri}, {Noguchi}, {Tarbell}, {Title},
  {Shine}, {Rosenberg}, {Hoffmann}, {Jurcevich}, {Kushner}, {Levay}, {Lites},
  {Elmore}, {Matsushita}, {Kawaguchi}, {Saito}, {Mikami}, {Hill}, \&
  {Owens}}]{2008SoPh..249..167T}
{Tsuneta}, S., {Ichimoto}, K., {Katsukawa}, Y., {Nagata}, S., {Otsubo}, M.,
  {Shimizu}, T., {Suematsu}, Y., {Nakagiri}, M., {Noguchi}, M., {Tarbell}, T.,
  {Title}, A., {Shine}, R., {Rosenberg}, W., {Hoffmann}, C., {Jurcevich}, B.,
  {Kushner}, G., {Levay}, M., {Lites}, B., {Elmore}, D., {Matsushita}, T.,
  {Kawaguchi}, N., {Saito}, H., {Mikami}, I., {Hill}, L.~D., \& {Owens}, J.~K.
  2008, \solphys, 249, 167

\bibitem[{{Van Doorsselaere} et~al.(2008){Van Doorsselaere}, {Nakariakov}, \&
  {Verwichte}}]{2008ApJ...676L..73V}
{Van Doorsselaere}, T., {Nakariakov}, V.~M., \& {Verwichte}, E. 2008, \apjl,
  676, L73

\bibitem[{{Zirker} et~al.(1994){Zirker}, {Engvold}, \&
  {Yi}}]{1994SoPh..150...81Z}
{Zirker}, J.~B., {Engvold}, O., \& {Yi}, Z. 1994, \solphys, 150, 81

\end{thebibliography}
\end{document}